\documentclass[doublecol]{epl2} 

\title{Prediction and inference in complex networks: a brief review and perspectives}

\author{Francisco A. Rodrigues}

\institute{                    
  \inst{1} Instituto de Ciências Matemáticas e de Computação, Universidade de São Paulo, São Carlos, São Paulo, Brazil\\
}

\abstract{
Inference and prediction are fundamental to the study of complex systems, where network data are often incomplete, inaccurate or obtained indirectly. In this paper, we review recent advances in network sampling and comparison, as well as in link prediction and network reconstruction from time series. We summarise key methodological developments and emerging approaches that integrate statistical and machine learning perspectives. We also outline promising research directions for enhancing the inference and prediction of complex networked systems.}
\begin{document}

\maketitle
The structure of complex systems can be represented by complex networks~\cite{Newman18:book}, which provide a unified framework for modelling phenomena ranging from interactions between individuals in online social platforms to biochemical processes~\cite{Costa11}. A central challenge in network research is to identify structural patterns that underlie system function and influence the evolution of dynamical processes~\cite{Boccaletti06}. Numerous studies have shown that features such as distances between nodes and connectivity heterogeneity can substantially influence phenomena including collective synchronisation~\cite{Rodrigues16} and epidemic spreading~\cite{Arruda18}.

To advance our understanding of complex systems, we need data that accurately captures their structure and dynamics. However, empirical network data is often incomplete or inaccurate, and is frequently limited by constraints in spatial and temporal resolution, which makes it difficult to observe all relevant interactions directly~\cite{Camps-Valls, Peel22, Kolaczyk14, Zou21}. In many cases, only a subset of the network can be observable~\cite{Leskovec}, or the available data consist of indirect measurements such as aggregated signals~\cite{Camps-Valls} and partial observations~\cite{Martinez}. As a result, inference and prediction methods are essential tools to extract meaningful information from limited data.

Inference, in broad terms, aims to extract information about a population based on observed samples~\cite{Heckathorn, Yousuf23, Kolaczyk14, Brugere18}. In ecological networks, for example, complete knowledge of interactions among species is seldom available~\cite{Fu19}; instead, analyses rely on samples obtained from diverse ecosystems~\cite{Mello19}. These samples allow researchers to infer global characteristics such as connectivity heterogeneity or the typical distance between species. Moreover, in many real-world settings, the network cannot be directly observed; only the signals it generates—such as electrical activity in the brain~\cite{Perovnik23}—are available. In such contexts, understanding system organisation requires inferring how nodes exchange information based on observed data~\cite{Bassett09, Alves23}.

Prediction addresses a complementary set of problems, aiming to anticipate future events, identify missing links, and estimate structural or dynamical changes in the system. For instance, predicting protein–protein interactions can help elucidate protein functions~\cite{Vazquez}, while in temporal networks~\cite{Holme12} it is possible to estimate which connections are likely to appear or change over time~\cite{Ahmed16, Zou21}. Predictive approaches also enable forecasting the behaviour of dynamical variables --- such as epidemic size or oscillator states --- by exploiting information encoded in network topology~\cite{Rodrigues25}.

Although inference and prediction pursue distinct objectives, they are closely intertwined. Both seek to extract meaningful information from incomplete or imperfect observations, and improvements in one often enhance the other: accurate inference strengthens predictive accuracy, while successful prediction provides evidence that can refine inferred models. Recent advances in network science and machine learning have further narrowed the gap between these two areas. Approaches such as graph neural networks~\cite{Corso24} and probabilistic graphical models~\cite{Koller09} simultaneously learn latent network representations and support robust predictions of node attributes, link formation, or system dynamics.

\begin{figure*}[!t]
\begin{center}
\includegraphics[width=1\linewidth]{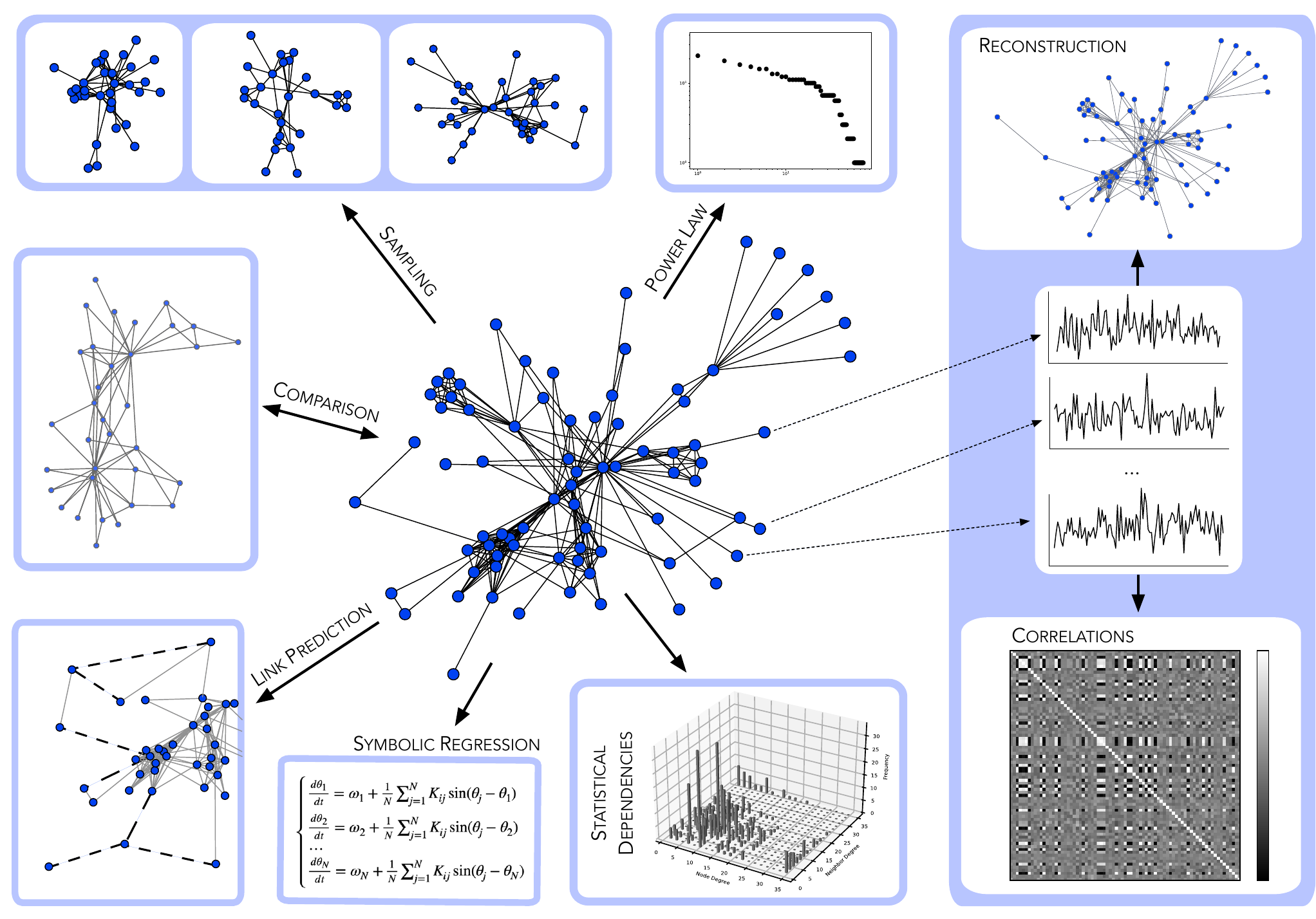}
\end{center}
\caption{Key challenges in network inference and prediction. Tasks such as sampling, network comparison and link prediction rely on samples of the network structure. In contrast, approaches such as symbolic regression, statistical dependency analysis and network reconstruction make use of metadata, which is often in the form of time series recorded from nodes and edges.
}
\label{Fig:inf-pred}
\end{figure*}

This work reviews key issues related to inference and prediction in complex networks (see figure~\ref{Fig:inf-pred}). We highlight significant advances and discuss methodological challenges, as well as outlining promising avenues for future research. Instead of providing an exhaustive review, we focus on key studies and emphasise conceptual developments with the greatest potential to advance the field.

\section{Network sampling}

Network sampling involves extracting a smaller subgraph from a large network while preserving its key structural properties. This enables analyses that are similar to those obtained from the full system. Most observed networks represent samples of larger, often inaccessible underlying systems. For example, although the human body contains between 80,000 and 400,000 proteins—implying approximately $1.6\times 10^5$ potential interactions—fewer than 150,000 have been mapped to date~\cite{Burke23, Smits16}. Similarly, our understanding of human mobility and social interactions relies on partial data, typically obtained from digital platforms or mobile devices~\cite{Gonzalez06}. Network sampling is therefore pervasive across disciplines—from ecology, in the construction of food webs, to economics, in the study of global trade patterns~\cite{Costa11}. Drawing reliable conclusions about the structure of the underlying system thus requires rigorous statistical inference.

Statistical inference provides a framework for extracting information about a population based on sample data~\cite{Wasserman13, Hu2013}. Because collecting complete information is often impractical or impossible, inference relies on estimating unknown parameters, constructing confidence intervals, and performing hypothesis tests using partial observations. These methods allow researchers to quantify uncertainty and ensure that conclusions are supported with statistically measurable reliability~\cite{Wasserman13}.

However, the foundations of traditional statistical inference --- particularly the assumption that samples are independent and identically distributed (i.i.d.)~\cite{Wasserman13} --- rarely hold in network contexts. Nodes are inherently dependent on one another through edges~\cite{Lee06}. In transportation networks, for example, disruptions in certain areas, such as traffic jams or flight delays, can spread across the entire system and affect many nodes. This structural dependence undermines the independence assumptions on which many classical inferential techniques are based, rendering them unsuitable for network data~\cite{Kolaczyk14}.

Network inference is further complicated by the lack of a universal sampling method~\cite{Yousuf23}. Simple random node sampling, though intuitive, tends to underestimate clustering and degree correlations because it ignores network dependencies~\cite{Kolaczyk14}. Random-walk sampling biases toward highly connected nodes, distorting degree distributions and global metrics. Snowball sampling effectively uncovers local communities but typically misses sparsely connected or isolated regions~\cite{Illenberger12, Heckathorn}. Edge-based sampling better captures adjacency relationships but disproportionately represents high-degree nodes, particularly in heterogeneous networks~\cite{Clauset03, Stumpf05, Heckathorn}. Therefore, each sampling technique carries its own structural biases, highlighting the difficulty of obtaining representative network samples.

The central challenge in network sampling, therefore, is to collect samples that preserve the structural properties of the original network~\cite{Yousuf23}. Sampled subgraphs should replicate key patterns such as degree distributions, correlation structures, and distance relationships. Determining the most appropriate sampling strategy, the minimum sample size required for accurate reconstruction and metrics for evaluating sampling properly remains an open problem.

Progress in this area may involve adapting established statistical tools. Methods based on maximum likelihood estimation and information criteria could guide model selection~\cite{VandeSchoot11}; Bayesian approaches can incorporate prior knowledge and quantify uncertainty in network estimates~\cite{VandeSchoot11}; and resampling techniques such as the bootstrap and jackknife can assess the robustness of inferences derived from partial data~\cite{Kunsch1989}. Although widely used in traditional statistics, these approaches have significant untapped potential for improving inference and prediction in complex networks, particularly when data are incomplete or biased.

These sampling limitations make it difficult to compare networks, since distortions introduced during data collection may mask or exaggerate structural differences. Therefore, understanding how to reliably compare networks requires a closer look at network comparison methods.

\section{Network comparison}

Comparing complex networks is essential both for identifying universal principles that govern diverse systems and for pinpointing structural features associated with specific functions. For example, examining the similarities and differences between biological, social and technological networks reveals recurring patterns that suggest shared evolutionary or functional mechanisms. Furthermore, comparing networks of the same type (e.g. social networks) sheds light on their unique properties, evolution and dynamics.
Despite its importance, network comparison remains a challenging task. Networks frequently differ in size, density, and organizational principles, complicating direct structural comparisons. Many key properties depend explicitly on the number of nodes $N$ and edges $M$. For example, in small-world networks the diameter scales as $\log \langle k \rangle / \log N$, where $\langle k \rangle = 2M/N$~\cite{Newman18}. Similarly, centrality measures such as betweenness and eigenvector centrality depend on shortest-path distances~\cite{Costa07}, which are themselves influenced by $N$ and $M$. As a result, direct comparisons across networks of different sizes or connectivities are often misleading unless appropriate normalization strategies are applied.

One approach to comparing networks is to analyze sampled subgraphs, provided that sampling methods adequately preserve the structural properties of the original networks~\cite{Lee06}. Another strategy involves learning vertex embeddings through machine-learning techniques~\cite{Xu}, which allow networks of varying sizes to be mapped into a common latent space. However, these embedding-based approaches often operate as black boxes~\cite{Xu}, limiting interpretability and potentially obscuring meaningful structural features.

Several alternative methodologies have been proposed. Genetic algorithms, for example, combine and weight multiple network features to construct similarity scores~\cite{Aliakbary15}, but they tend to be computationally expensive and lack strong theoretical guarantees. Network taxonomies based on structural patterns such as community organization offer another possibility~\cite{Onnela12}, though their reliability depends heavily on the accuracy of community detection methods—a significant drawback given that many empirical networks do not exhibit clear modular structure.

Additional tools for network comparison include spectral methods~\cite{Domenico16}, statistical distributions analysis~\cite{Schieber}, ensemble graph distances~\cite{Hartle20}, and information-theoretic metrics~\cite{Bagrow19}. Despite their usefulness, these approaches still lack a unified theoretical framework grounded in statistical inference. Developing such a foundation remains an open challenge and is essential for advancing the systematic comparison of complex networks.

More recently, renormalization-inspired techniques have been explored as a promising alternative~\cite{Villegas23, Gabrielli25}. Adapting renormalization group concepts to networks, however, presents substantial theoretical and practical difficulties: nodes and subgraphs often display heterogeneous statistical behaviours; most networks lack geometric coordinates; and the lattice-like symmetries that underpin renormalization in physical systems are generally absent~\cite{Gabrielli25}. Nonetheless, despite these challenges, renormalization-based approaches represent an exciting direction with the potential to yield transformative insights into the multiscale structure of complex networks.

\section{Link prediction}

Link prediction plays a central role in network science~\cite{lu2011link, Martinez, Xie, Menand, Zhou21}, as it enables the identification of unobserved connections and the forecasting of future interactions in temporal networks~\cite{Zou21}. This task underlies many recommendation systems and is widely applied in social network analysis to anticipate potential friendships, uncover community structures, and examine patterns of information diffusion~\cite{Martinez}.

In biological systems, link prediction offers valuable insights into complex interaction networks. For example, in protein–protein interaction networks, structural and biological features can be combined to predict missing interactions, thereby advancing our understanding of cellular organization and function~\cite{Kovacs19}. Similarly, in drug–drug interaction networks, predicting potential synergies or antagonisms between compounds can accelerate drug discovery~\cite{Abbas21}. In social platforms, link prediction supports user engagement by suggesting new connections, while in criminal or corruption networks it can reveal hidden actors and uncover suspicious relationships~\cite{Martinez}.

A wide range of methodological frameworks has been developed to address the link prediction problem~\cite{Martinez}. Similarity-based approaches measure the relatedness between nodes using metrics such as Jaccard or cosine similarity~\cite{Martinez}. Embedding-based methods~\cite{Wang22}, including Node2Vec and DeepWalk, map nodes into a latent space in which distances reflect the likelihood of link formation~\cite{Wu22}. Machine-learning methods formulate link prediction as a binary classification task~\cite{Rodrigues23}, employing models such as neural networks or ensemble methods to infer potential connections~\cite{Martinez}. Probabilistic models provide yet another avenue by estimating the likelihood of connections based on noisy or incomplete observations~\cite{Newman18}. Each of these approaches has its own strengths and limitations, which depend on network characteristics, data availability, and prediction objectives.

In addition to the structural complexity of networks, external factors play a significant role in shaping interactions, further complicating link prediction. In social systems, for example, user behaviour is highly dynamic and influenced by evolving trends and external events, leading to fluctuations that are difficult to model. To address these challenges, researchers increasingly rely on probabilistic models, advanced deep learning architectures, and the integration of heterogeneous data sources~\cite{Martinez}. Although hybrid approaches that combine these techniques show considerable promise, further work is required to refine existing methods, enhance interpretability, and improve predictive performance in real-world settings~\cite{Martinez, Wu22}.

\section{Network reconstruction and statistical dependencies}

Time series recorded from complex systems can be used for two primary purposes: (i) topological inference and reconstruction~\cite{Camps-Valls}, and (ii) influence mapping~\cite{Sporns13, Bassett09}. In the first case, researchers use simulated dynamics as benchmarks to design and validate algorithms capable of inferring the underlying network from observed signals~\cite{Camps-Valls, Peixoto19, Young20}. A dynamical process is simulated on a known network, the resulting time series are recorded, and the underlying topology is subsequently reconstructed using statistical inference or machine-learning methods. Comparing the reconstructed network with the original one provides a quantitative measure of algorithmic accuracy.

In influence mapping, by contrast, the objective is to identify statistical relationships from time series independently of the physical connectivity of the system. Here, functional correlations between signals do not necessarily correspond to physical edges but can nevertheless reveal meaningful patterns of interaction~\cite{Peron11, Alves23}. For example, brain studies use functional networks to compare patients with disorders such as epilepsy or autism against healthy controls~\cite{Alves23, Basset17}. In climate science, the goal is to understand how large-scale oscillators such as El Niño influence regional weather patterns~\cite{Yamasaki08}.

A wide range of methods exists for reconstructing networks and quantifying statistical dependencies from time series. These include correlation-based measures (e.g., Pearson and Spearman correlations) and causal inference methods~\cite{Camps-Valls}. However, the limitations of correlation are well known: correlation does not imply causation, and spurious correlations can arise when nodes share a common driver~\cite{Peel22}. As a result, correlation-based methods must be applied with caution. Despite these issues, they remain useful in specific contexts --- for instance, in the comparison of functional brain networks for diagnostic purposes~\cite{Bassett09, Perovnik23}.

Causal inference methods, although powerful, introduce additional challenges~\cite{Camps-Valls}. They require strong modelling assumptions, high-quality time-series data, and sophisticated algorithms to detect true causal relationships. Common techniques include Granger causality, mutual information, and transfer entropy~\cite{Camps-Valls}, each with distinct advantages and limitations.

From a statistical perspective, many correlation-based approaches can be interpreted as instances of Bayesian modelling under non-informative priors. Bayesian inference methods can also reconstruct networks directly from time series~\cite{Peel22}, though they typically require assumptions about model structure. Their major drawback is computational cost, which restricts their application to relatively small networks~\cite{Newman18:book}. Nonetheless, Bayesian approaches remain promising due to their principled nature and robustness~\cite{Newman18, Peixoto19}.

Although substantial progress has been made, the reconstruction and quantification of dependencies in functional networks remains an open problem. This area of research holds significant potential and is likely to play a key role in advancing the analysis of complex systems characterised by time-series data—from climate and biological systems to financial and technological networks.

\section{Prediction of dynamic processes}

It is well established that network structure strongly influences dynamical processes; for instance, the critical coupling required for the synchronization of Kuramoto oscillators depends on the moments of the degree distribution~\cite{Boccaletti06}. This relationship allows structural information to be used for predicting dynamical variables by analysing patterns of connectivity in the underlying network. For example, Rodrigues et al.~\cite{Rodrigues25} showed that it is possible to predict epidemic size or the state of a Kuramoto oscillator based solely on network connection patterns.

Despite these promising initial results, the integration of machine learning and multivariate statistical methods to study structure–dynamics relationships remains in its early stages, with only a limited number of studies available (e.g.~\cite{Arruda13, Pineda23, Rodrigues25}). Extending these approaches to other classes of dynamical processes—such as cooperation, cascade failures, or rumour spreading~\cite{Boccaletti06}—represents a natural next step. Additionally, techniques such as symbolic regression~\cite{Brum25} offer a complementary avenue for uncovering the influence of network structure on dynamics by generating interpretable equations that describe the evolution of complex systems and explicitly capture structure–dynamics dependencies~\cite{Menezes14}.

\section{Statistical dependencies}

Quantifying statistical dependencies in network structure --- such as degree correlations --- is essential for understanding a wide range of dynamical processes. 
The conditional probability $P(K \mid K')$, where $K$ and $K'$ denote the degrees of adjacent nodes, characterises network assortativity and influences phenomena such as synchronisation and epidemic spreading~\cite{Rodrigues16, Arruda18, Newman18}. 
However, estimating $P(K \mid K')$ in empirical networks remains challenging due to limited sample sizes, sampling biases, sparsity, and long-range dependencies that interact with other structural properties, including clustering and community organisation~\cite{Boccaletti06}. 
These difficulties highlight the need for more robust methods to quantify degree correlations.

One promising alternative is the use of \textit{copulas}~\cite{Nelsen06}, which model dependency structures independently of marginal distributions and can capture complex, nonlinear relationships --- properties frequently exhibited by real-world networks but still largely unexplored in this context.

Importantly, conditional dependencies in networks need not be restricted to node degree. 
More general conditional distributions $P(X \mid Y)$, where $X$ and $Y$ denote topological or dynamical features, can be employed to study a broader range of structural and functional relationships. 
Such approaches naturally extend beyond pairwise dependencies and can characterise higher-order, multivariate interaction patterns within networks~\cite{Ferraz24}. 
The analysis of these structural and dynamical dependencies defines a vast and still underdeveloped research frontier with significant theoretical and practical potential.

\section{Power laws}

Another longstanding challenge concerns the identification of scale-free behaviour in empirical networks. 
Although many real-world networks appear to follow power-law degree distributions, this behaviour often holds only over limited ranges~\cite{Broido19}. 
Data quality presents further complications: empirical social and biological networks frequently contain sampling biases, particularly underrepresenting nodes with very low or very high degrees. 
Moreover, different sampling strategies can distort the tail behaviour of degree distributions, potentially leading to incorrect conclusions about the presence of power laws --- or even causing purely random networks to appear scale-free~\cite{Clauset03, Stumpf05}. 
Network size also plays a critical role, as small or fragmented datasets can significantly affect inference accuracy~\cite{Kolaczyk14}.

\section{Other perspectives}

Beyond the challenges already discussed, numerous open problems remain in network inference and prediction. 
Community detection~\cite{Fortunato16} and pattern recognition~\cite{Costa09, Costa09EPL} continue to be active research areas, where both exact and approximate inference methods, Bayesian model-selection techniques~\cite{Peixoto15}, and hierarchical modelling frameworks~\cite{Peixoto14} offer principled statistical approaches for identifying latent structures and rigorously evaluating competing hypotheses under uncertainty and incomplete data.

Finally, many inference and prediction studies focus on simplified network representations, often neglecting important structural features such as temporal dynamics~\cite{Holme12}, multilayer or multiplex organisation~\cite{Cozzo}, higher-order interactions~\cite{Ferraz24}, and directed or weighted connections~\cite{Boccaletti06}. 
Addressing these complexities requires continued development of statistical and machine-learning techniques capable of capturing the full richness of real-world networks.

\section{Conclusion}

Inference and prediction are fundamental to the study of complex networks, especially given that real-world network data are frequently incomplete, noisy, or derived from indirect measurements such as time-series signals~\cite{Kolaczyk14}. Consequently, uncertainty pervades all stages of network analysis, exerting significant influence on diverse fields including social networks, climate science, neuroscience, and biological systems~\cite{Costa11}. In this work, we discuss several challenges associated with uncertainty in networks, including issues related to sampling, influence quantification, network reconstruction, and structural comparison. Furthermore, we emphasise the pivotal role of multivariate statistical methods and machine learning techniques in identifying pertinent patterns that elucidate the interplay between structure and dynamics in complex systems.

Despite the substantial progress that has been made, several challenges remain to be addressed. The development of a principled framework based on statistical theory is essential, especially one that incorporates Bayesian methodologies that can update inferences when new data becomes available~\cite{Peel22}. Contemporary machine learning methodologies, including graph neural networks~\cite{Corso24}, copulas~\cite{Nelsen06} and causal inference techniques~\cite{Camps-Valls}, offer powerful tools for managing nonlinear relationships, temporal evolution~\cite{Holme12} and higher-order dependencies~\cite{Ferraz24}.

These challenges are complex, but must be addressed to drive progress in neuroscience, ecology, social systems and climate modelling applications. Advances in these areas could transform the way we analyse and interpret complex systems, offering a more comprehensive and integrated understanding of their structure and dynamics.

\acknowledgments
Francisco Rodrigues acknowledges financial support from CNPq (grants 308162/2023-4 and 408389/2024-9) and FAPESP (grants 24/02322-0, 20/09835-1, and 13/07375-0) for his research.

\bibliographystyle{eplbib}
\bibliography{references}

\begin{thebibliography}{10}
\expandafter\ifx\csname url\endcsname\relax\def\url#1{\texttt{#1}}\fi

\bibitem{Newman18:book}
\Name{Newman M.} \Book{Networks} ({Oxford University Press}) 2018.

\bibitem{Costa11}
\Name{Costa L. d.~F., Oliveira~Jr O.~N., Travieso G., Rodrigues F.~A.,
  Villas~Boas P.~R., Antiqueira L., Viana M.~P. \and Correa~Rocha L.~E.}
  \REVIEW{Advances in Physics}{60}{2011}{329}.

\bibitem{Boccaletti06}
\Name{Boccaletti S., Latora V., Moreno Y., Chavez M. \and Hwang D.-U.}
  \REVIEW{Physics Reports}{424}{2006}{175}.

\bibitem{Rodrigues16}
\Name{Rodrigues F.~A., Peron T. K.~D., Ji P. \and Kurths J.} \REVIEW{Physics
  Reports}{610}{2016}{1}.

\bibitem{Arruda18}
\Name{de~Arruda G.~F., Rodrigues F.~A. \and Moreno Y.} \REVIEW{Physics
  Reports}{756}{2018}{1}.

\bibitem{Camps-Valls}
\Name{Camps-Valls G., Gerhardus A., Ninad U., Varando G., Martius G.,
  Balaguer-Ballester E., Vinuesa R., Diaz E., Zanna L. \and Runge J.}
  \REVIEW{Physics Reports}{1044}{2023}{1}.

\bibitem{Peel22}
\Name{Peel L., Peixoto T.~P. \and De~Domenico M.} \REVIEW{Nature
  Communications}{13}{2022}{6794}.

\bibitem{Kolaczyk14}
\Name{Kolaczyk E.~D. \and Cs{\'a}rdi G.} \Book{Statistical analysis of network
  data with R} Vol.~65 (Springer) 2014.

\bibitem{Zou21}
\Name{Zou L., Zhan X.-X., Sun J., Hanjalic A. \and Wang H.} \REVIEW{IEEE
  Transactions on Network Science and Engineering}{9}{2021}{1215}.

\bibitem{Leskovec}
\Name{Leskovec J. \and Faloutsos C.} \Book{Sampling from large graphs} in proc.
  of \Book{Proceedings of the 12th ACM SIGKDD international conference on
  Knowledge discovery and data mining} 2006 pp. 631--636.

\bibitem{Martinez}
\Name{Mart{\'\i}nez V., Berzal F. \and Cubero J.-C.} \REVIEW{ACM computing
  surveys (CSUR)}{49}{2016}{1}.

\bibitem{Heckathorn}
\Name{Heckathorn D.~D. \and Cameron C.~J.} \REVIEW{Annual review of
  sociology}{43}{2017}{101}.

\bibitem{Yousuf23}
\Name{Yousuf M.~I., Anwer I. \and Anwar R.} \REVIEW{Social Network Analysis and
  Mining}{13}{2023}{66}.

\bibitem{Brugere18}
\Name{Brugere I., Gallagher B. \and Berger-Wolf T.~Y.} \REVIEW{ACM Computing
  Surveys}{51}{2018}{1}.

\bibitem{Fu19}
\Name{Fu X., Seo E., Clarke J. \and Hutchinson R.~A.} \REVIEW{IEEE Transactions
  on Knowledge and Data Engineering}{33}{2019}{3117}.

\bibitem{Mello19}
\Name{Mello M.~A., Felix G.~M., Pinheiro R.~B., Muylaert R.~L., Geiselman C.,
  Santana S.~E., Tschapka M., Lotfi N., Rodrigues F.~A. \and Stevens R.~D.}
  \REVIEW{Nature Ecology \& Evolution}{3}{2019}{1525}.

\bibitem{Perovnik23}
\Name{Perovnik M., Rus T., Schindlbeck K.~A. \and Eidelberg D.} \REVIEW{Nature
  Reviews Neurology}{19}{2023}{73}.

\bibitem{Bassett09}
\Name{Bassett D.~S. \and Bullmore E.~T.} \REVIEW{Current Opinion in
  Neurology}{22}{2009}{340}.

\bibitem{Alves23}
\Name{Alves C.~L., Toutain T. G. d.~O., de~Carvalho~Aguiar P., Pineda A.~M.,
  Roster K., Thielemann C., Porto J. A.~M. \and Rodrigues F.~A.}
  \REVIEW{Scientific Reports}{13}{2023}{8072}.

\bibitem{Vazquez}
\Name{Vazquez A., Flammini A., Maritan A. \and Vespignani A.} \REVIEW{Nature
  Biotechnology}{21}{2003}{697}.

\bibitem{Holme12}
\Name{Holme P. \and Saram{\"a}ki J.} \REVIEW{Physics Reports}{519}{2012}{97}.

\bibitem{Ahmed16}
\Name{Ahmed N.~M., Chen L., Wang Y., Li B., Li Y. \and Liu W.}
  \REVIEW{Information Sciences}{374}{2016}{1}.

\bibitem{Rodrigues25}
\Name{Rodrigues F.~A., Peron T., Connaughton C., Kurths J. \and Moreno Y.}
  \REVIEW{Proceedings of the Royal Society A}{481}{2025}{20240435}.

\bibitem{Corso24}
\Name{Corso G., Stark H., Jegelka S., Jaakkola T. \and Barzilay R.}
  \REVIEW{Nature Reviews Methods Primers}{4}{2024}{17}.

\bibitem{Koller09}
\Name{Koller D. \and Friedman N.} \Book{Probabilistic graphical models:
  principles and techniques} (MIT press) 2009.

\bibitem{Burke23}
\Name{Burke D.~F., Bryant P., Barrio-Hernandez I., Memon D., Pozzati G., Shenoy
  A., Zhu W., Dunham A.~S., Albanese P., Keller A. \etal} \REVIEW{Nature
  Structural \& Molecular Biology}{30}{2023}{216}.

\bibitem{Smits16}
\Name{Smits A.~H. \and Vermeulen M.} \REVIEW{Trends in
  Biotechnology}{34}{2016}{825}.

\bibitem{Gonzalez06}
\Name{Gonz{\'a}lez M.~C., Lind P.~G. \and Herrmann H.~J.} \REVIEW{Physical
  Review Letters}{96}{2006}{088702}.

\bibitem{Wasserman13}
\Name{Wasserman L.} \Book{All of statistics: a concise course in statistical
  inference} (Springer Science \& Business Media) 2013.

\bibitem{Hu2013}
\Name{Hu P. \and Lau W.~C.} \REVIEW{Preprint ArXiv:1308.5865}{}{2013}{}.

\bibitem{Lee06}
\Name{Lee S.~H., Kim P.-J. \and Jeong H.} \REVIEW{Physical Review
  E}{73}{2006}{016102}.

\bibitem{Illenberger12}
\Name{Illenberger J. \and Fl{\"o}tter{\"o}d G.} \REVIEW{Social
  Networks}{34}{2012}{701}.

\bibitem{Clauset03}
\Name{Clauset A. \and Moore C.} \REVIEW{arXiv preprint
  cond-mat/0312674}{}{2003}{}.

\bibitem{Stumpf05}
\Name{Stumpf M.~P., Wiuf C. \and May R.~M.} \REVIEW{Proceedings of the National
  Academy of Sciences}{102}{2005}{4221}.

\bibitem{VandeSchoot11}
\Name{Van~de Schoot R., Depaoli S., King R., Kramer B., M{\"a}rtens K., Tadesse
  M.~G., Vannucci M., Gelman A., Veen D., Willemsen J. \etal} \REVIEW{Nature
  Reviews Methods Primers}{1}{2021}{1}.

\bibitem{Kunsch1989}
\Name{K{\"u}nsch H.~R.} \REVIEW{The annals of Statistics}{}{1989}{1217}.

\bibitem{Newman18}
\Name{Newman M.~E.} \REVIEW{Nature Physics}{14}{2018}{542}.

\bibitem{Costa07}
\Name{Costa L. d.~F., Rodrigues F.~A., Travieso G. \and Villas~Boas P.~R.}
  \REVIEW{Advances in Physics}{56}{2007}{167}.

\bibitem{Xu}
\Name{Xu M.} \REVIEW{SIAM Review}{63}{2021}{825}.

\bibitem{Aliakbary15}
\Name{Aliakbary S., Motallebi S., Rashidian S., Habibi J. \and Movaghar A.}
  \REVIEW{Chaos: An Interdisciplinary Journal of Nonlinear
  Science}{25}{2015}{}.

\bibitem{Onnela12}
\Name{Onnela J.-P., Fenn D.~J., Reid S., Porter M.~A., Mucha P.~J., Fricker
  M.~D. \and Jones N.~S.} \REVIEW{Physical Review E}{86}{2012}{036104}.

\bibitem{Domenico16}
\Name{De~Domenico M. \and Biamonte J.} \REVIEW{Physical Review
  X}{6}{2016}{041062}.

\bibitem{Schieber}
\Name{Schieber T.~A., Carpi L., D{\'\i}az-Guilera A., Pardalos P.~M., Masoller
  C. \and Ravetti M.~G.} \REVIEW{Nature Communications}{8}{2017}{13928}.

\bibitem{Hartle20}
\Name{Hartle H., Klein B., McCabe S., Daniels A., St-Onge G., Murphy C. \and
  H{\'e}bert-Dufresne L.} \REVIEW{Proceedings of the Royal Society
  A}{476}{2020}{20190744}.

\bibitem{Bagrow19}
\Name{Bagrow J.~P. \and Bollt E.~M.} \REVIEW{Applied Network
  Science}{4}{2019}{1}.

\bibitem{Villegas23}
\Name{Villegas P., Gili T., Caldarelli G. \and Gabrielli A.} \REVIEW{Nature
  Physics}{19}{2023}{445}.

\bibitem{Gabrielli25}
\Name{Gabrielli A., Garlaschelli D., Patil S.~P. \and Serrano M.~{\'A}.}
  \REVIEW{Nature Reviews Physics}{}{2025}{1}.

\bibitem{lu2011link}
\Name{L{\"u} L. \and Zhou T.} \REVIEW{Physica A: statistical mechanics and its
  applications}{390}{2011}{1150}.

\bibitem{Xie}
\Name{He X., Ghasemian A., Lee E., Schwarze A.~C., Clauset A. \and Mucha P.~J.}
  \REVIEW{Plos one}{19}{2024}{e0306883}.

\bibitem{Menand}
\Name{Menand N. \and Seshadhri C.} \REVIEW{Proceedings of the National Academy
  of Sciences}{121}{2024}{e2312527121}.

\bibitem{Zhou21}
\Name{Zhou T.} \REVIEW{Iscience}{24}{2021}{}.

\bibitem{Kovacs19}
\Name{Kov{\'a}cs I.~A., Luck K., Spirohn K., Wang Y., Pollis C., Schlabach S.,
  Bian W., Kim D.-K., Kishore N., Hao T. \etal} \REVIEW{Nature
  Communications}{10}{2019}{1240}.

\bibitem{Abbas21}
\Name{Abbas K., Abbasi A., Dong S., Niu L., Yu L., Chen B., Cai S.-M. \and
  Hasan Q.} \REVIEW{BMC Bioinformatics}{22}{2021}{1}.

\bibitem{Wang22}
\Name{Wang Z., Zhan X.-X., Liu C. \and Zhang Z.-K.}
  \REVIEW{Iscience}{25}{2022}{}.

\bibitem{Wu22}
\Name{Wu H., Song C., Ge Y. \and Ge T.} \REVIEW{Data Science and
  Engineering}{7}{2022}{253}.

\bibitem{Rodrigues23}
\Name{Rodrigues F.~A.} \REVIEW{Europhysics Letters}{144}{2023}{22001}.

\bibitem{Sporns13}
\Name{Sporns O.} \REVIEW{Dialogues in clinical neuroscience}{15}{2013}{247}.

\bibitem{Peixoto19}
\Name{Peixoto T.~P.} \REVIEW{Physical Review Letters}{123}{2019}{128301}.

\bibitem{Young20}
\Name{Young J.-G., Cantwell G.~T. \and Newman M.} \REVIEW{Journal of Complex
  Networks}{8}{2020}{cnaa046}.

\bibitem{Peron11}
\Name{Peron T.~D. \and Rodrigues F.~A.} \REVIEW{Europhysics
  Letters}{96}{2011}{48004}.

\bibitem{Basset17}
\Name{Bassett D.~S. \and Sporns O.} \REVIEW{Nature
  Neuroscience}{20}{2017}{353}.

\bibitem{Yamasaki08}
\Name{Yamasaki K., Gozolchiani A. \and Havlin S.} \REVIEW{Physical Review
  Letters}{100}{2008}{228501}.

\bibitem{Arruda13}
\Name{de~Arruda G.~F., Dal’Maso~Peron T.~K., de~Andrade M.~G., Achcar J.~A.
  \and Rodrigues F.~A.} \REVIEW{Journal of Statistical
  Physics}{152}{2013}{519}.

\bibitem{Pineda23}
\Name{Pineda A.~M., Kent P., Connaughton C. \and Rodrigues F.~A.}
  \REVIEW{Journal of Statistical Mechanics: Theory and
  Experiment}{2023}{2023}{123402}.

\bibitem{Brum25}
\Name{Brum B.~R., Lober L., Previdelli I. \and Rodrigues F.~A.}
  \REVIEW{Preprint ArXiv:2508.20257}{}{2025}{}.

\bibitem{Menezes14}
\Name{Menezes T. \and Roth C.} \REVIEW{Scientific reports}{4}{2014}{6284}.

\bibitem{Nelsen06}
\Name{Nelsen R.~B.} \Book{An introduction to copulas} (Springer) 2006.

\bibitem{Ferraz24}
\Name{Ferraz~de Arruda G., Aleta A. \and Moreno Y.} \REVIEW{Nature Reviews
  Physics}{6}{2024}{468}.

\bibitem{Broido19}
\Name{Broido A.~D. \and Clauset A.} \REVIEW{Nature
  Communications}{10}{2019}{1017}.

\bibitem{Fortunato16}
\Name{Fortunato S. \and Hric D.} \REVIEW{Physics Reports}{659}{2016}{1}.

\bibitem{Costa09}
\Name{Costa L. d.~F. \and Rodrigues F.~A.} \REVIEW{Europhysics
  Letters}{85}{2009}{48001}.

\bibitem{Costa09EPL}
\Name{Costa L. d.~F., Rodrigues F.~A., Hilgetag C.~C. \and Kaiser M.}
  \REVIEW{Europhysics Letters}{87}{2009}{18008}.

\bibitem{Peixoto15}
\Name{Peixoto T.~P.} \REVIEW{Physical Review X}{5}{2015}{011033}.

\bibitem{Peixoto14}
\Name{Peixoto T.~P.} \REVIEW{Physical Review X}{4}{2014}{011047}.

\bibitem{Cozzo}
\Name{Cozzo E., De~Arruda G.~F., Rodrigues F.~A. \and Moreno Y.}
  \Book{Multiplex networks: basic formalism and structural properties} Vol.~2
  (Springer) 2018.

\end{thebibliography}

\end{document}